# The role of electronic bandstructure shape in improving the thermoelectric power factor of complex materials

Patrizio Graziosi[1*] and Neophytos Neophytou[2]

[1] Institute of Nanostructured Materials, CNR, Bologna, Italy
[2] School of Engineering, University of Warwick, Coventry, CV4 7AL, UK

*patrizio.graziosi@gmail.com

## Abstract

The large variety of complex electronic structure materials and their alloys, offer highly promising directions for improvements in thermoelectric (TE) power factors (PF). Their electronic structure contains rich features, referred to as 'surface complexity', one of them being the highly anisotropic warped energy surface shapes with elongated features and threads in some cases. In this work we use Boltzmann transport simulations to quantify the influence that the shape of the electronic structure energy surfaces has on the PF. Using both analytical ellipsoidal bands, as well as realistic bands from the group of half-Heuslers, we show that band shape complexity alone can offer an advantage to the PF of ~3× in realistic cases. The presence of anisotropic scattering mechanisms such as ionized impurity or polar optical phonon scattering, however, can reduce these improvements by up to ~50%. We show that expressions based on the simple ratio of the density-of-states to the conductivity effective masses, $m_{DOS}/m_C$, together with the number of valleys, can capture the anisotropy shape with a moderate to high degree of correlation. For this, we use a convenient way to extract these masses by mapping the complex bandstructures of materials to parabolic electronic structures, without the need for Boltzmann transport codes. Despite the fact that the PF depends on many parameters, information about the benefits of the band shape alone, would be very useful for identifying and understanding the performance of novel thermoelectric materials.

# Introduction

Thermoelectric generators (TEGs) are solid-state devices able to convert the heat flow arising from temperature gradients directly into electricity [1, 2]. They have the potential to offer a sustainable path for power harvesting as well as cooling across a variety of industries, including medical, wearable electronics, building monitoring, the internet of things, refrigeration, thermal management, space missions, transportation, and many more. However, large scale exploitation has been limited by the high prices, toxicity, scarcity, and the low efficiencies of the prominent thermoelectric (TE) materials.

The TE performance is quantified by the $ZT$ figure of merit as $\sigma S^2 T/(\kappa_e + \kappa_l)$, where $\sigma$ is the electrical conductivity, $S$ is the Seebeck coefficient, $T$ is the absolute temperature, and $\kappa_e$ and $\kappa_l$ are the electronic and lattice parts of the thermal conductivity, respectively. The product $\sigma S^2$ in the numerator of $ZT$ is called the power factor (PF). Over the last two decades, progress on TE materials has been rapidly expanding with the synthesis of a myriad of new materials and their alloys. While the thermal conductivity was targeted by nanostructuring which can effectively reduce $\kappa_l$, the PF is targeted by the complex features that new materials exhibit in their electronic structure [1, 2]. These electronic structures typically contain rich features such many valleys [3], warped features [4], resonant states [3, 5, 6, 7], narrow lower-dimensionality threads in 3D [8], topological states [9, 10], and more (some are included in descriptors and fitness functions used in machine learning studies [11, 12, 13, 14, 15]). Many of these features have been studied in detail, a research direction referred to as 'bandstructure engineering', which targets novel transport features and PF enhancement [16, 17, 18, 19, 20, 21]. Indeed, some of the best TE materials such as PbTe, $Bi_2Te_3$, and PbSe have multiple valleys and highly anisotropic band features with different masses along different directions [12, 22, 23, 24, 25, 26].

Specifically, the influence of the shape of the electronic structure, with its band warping, elongated features, and high band anisotropy on the PF, has been a subject of investigations over the last years [8, 27, 28, 29, 30, 31]. The rationale is that compared to a fully isotropic band with the same density of states (DOS), an ellipsoidal anisotropic band can offer higher conductivity in a certain direction. Indeed, the conductivity effective mass, $m_C$, of an ellipsoidal band is given by the harmonic ratio of the ellipsoidal masses as $m_c =$ $3(\frac{1}{m_l} + \frac{2}{m_t})^{-1}$, whereas the density of states effective mass, $m_{DOS}$, which influences the Seebeck coefficient, is given by the geometric average of the ellipsoidal masses as $m_{DOS} = (m_l m_t^2)^{1/3}$. Anisotropy can decouple these two to achieve high conductivity and Seebeck coefficients [29], which typically is difficult to reach, given the adverse interdependence of the two quantities. This can make ellipsoidal bands beneficial to the PF, with variants of the ratio of $m_{DOS}/m_C$ suggested in the past as a descriptor for high PF TE materials [27]. On the other hand, a realistic bandstructure will include many ellipsoids along the high symmetry axes, which will average the overall bandstructure into something more isotropic. In addition, transport simulations using theoretical models for warped bands have also demonstrated strong influence on the thermoelectric coefficients [31]. Another way in which anisotropy appears, is that in many promising TE materials such as as GeTe and SnTe, there can be corrugated energy surfaces with large deviations from ellipsoidal shapes, even for relatively low energies and carrier concentrations such [4, 28, 32]. This feature can also be beneficial for TEs as a corrugated 3D surface encloses larger DOS, which improves the Seebeck coefficient. In fact, it was shown that GeTe and SnTe with corrugated energy surfaces, have a significant PF advantage compared to the ellipsoidal PbTe and PbSe due to corrugation alone [28]. Other works investigate materials with pudding-mold-type bands with partially flat regions, which largely increase anisotropy as well, and report large Seebeck coefficients [33, 34].

Prior computational works have studied the effects of band anisotropy to the PF using Boltzmann transport, but by using the simple constant relaxation time approximation (CRTA) for scattering and have not quantified the magnitude of this effect reliably. The effects of detailed and realistic energy-dependent scattering mechanisms, the effects of intra- versus inter-valley scattering, the consideration of anisotropic (momentum dependent) scattering mechanisms such as ionized impurity scattering and polar optical phonon scattering, are still unknown, despite being dominant in TE materials. Moreover, there are different methods on how the masses that form the descriptor $m_{DOS}/m_C$ are extracted, to provide a measure of anisotropy. Prior works used the mobility and Seebeck coefficient obtained from Boltzmann transport calculations under the CRTA to extract them [27]. Here, we revisit the topic of the influence of the shape of the energy surfaces on the thermoelectric PF. We take a different approach to obtain the effective masses without the use of Boltzmann transport software, we quantify the influence of the anisotropy on the PF, we present a thorough investigation of how different scattering processes affect the suggested PF benefits, and examine this effect for the half-Heusler (HH) material family by using a group of 14 n-type and 14 p-type HHs.

The paper is organized as follows: we start by quantifying the impact of anisotropy on the PF considering a single ellipsoidal band. Then, we perform an analysis of the impact of anisotropy on the PF by considering six ellipsoidal bands and a series of different scattering mechanisms that are typically encountered in TE materials. Further, we perform the same study, but for realistic DFT extracted HH bandstructures, and finally we conclude.

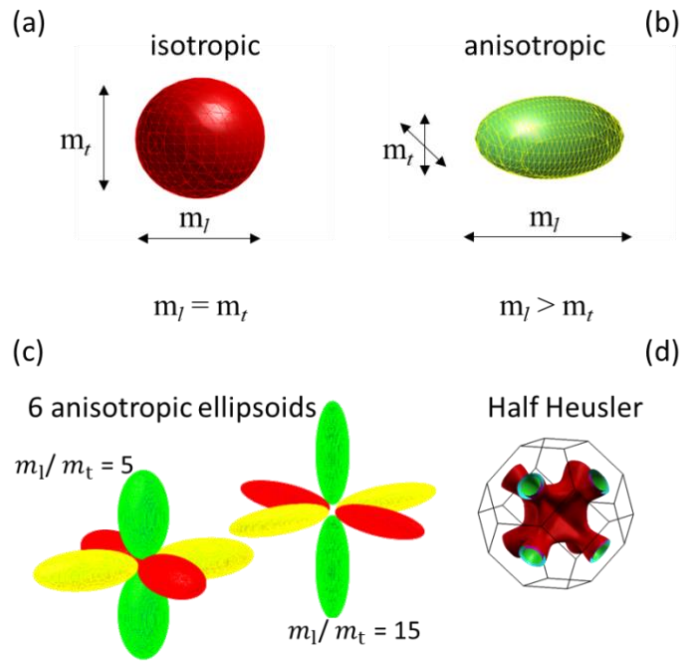

**Figure 1**: (a) Constant energy surfaces for: (a) an isotropic parabolic band (sphere); (b) an anisotropic parabolic band (ellipsoid); (c) for a bandstructure with 6 valleys placed at the middle of the Γ-X line (conjugated valleys are depicted with the same color); and (d) of one of the three valence bands of TiCoSb, as representative of complex bandstructures in half-Heusler alloys.
4

## Anisotropy considerations for a single ellipsoidal band

The general question we ask in this work is, if compared to a fully isotropic spherical energy surface band, a bandstructure of the same DOS but composing of anisotropic and/or highly warped bands will provide higher PFs, and how much higher, due to the band shape alone. Figure 1 shows examples of these distinctive features between an isotropic band (Fig. 1a), a bandstructure with anisotropic ellipsoidal energy surfaces (Fig. 1b and 1c), and more complex, realistic warped energy surfaces (Fig. 1d). Such electronic structures provide states with both higher and lower velocities, scattering rates, and local density of states, compared to a spherical band in which all states have the same attributes on a given energy surface. Note that we are not investigating the TE performance of anisotropic crystal, low-symmetry materials, i.e. hexagonal $Mg_3Sb_2$, for which the *c*-axis would perform differently compared to the *a*- and *b*-axes [35]. What we focus on is the surface complexity and anisotropy even in isotropic crystal cubic materials of high symmetry (like HHs). This is also relevant for polycrystals, where the properties are averaged along different directions and we again expect the material to be overall more isotropic.

We start by investigating the effect of anisotropy of a single ellipsoidal band (Fig. 1b), in comparison to a fully isotropic band (Fig. 1a). This will provide a first order understanding for the band shape and the effects of different scattering mechanisms, before we consider realistic bands. We use parabolic effective mass theory to describe the bands, thought we are fully aware that such treatment cannot in general capture band corrugation or warping. However, this simplified treatment can allow for clarity in understanding and ease in computation. We describe below how appropriately extracted mass values can be defined for complex bands, and that they can adequately describe the effects of warping and corrugation on transport for a general material. Such simple treatment on the effective mass description level can then be used easily in materials screening and machine learning studies, for example.

We compute electronic transport using the linearized Boltzmann transport equation (BTE). All simulations are performed at the temperature of $T = 300$ K. Within the BTE, the TE coefficients are given by: [36, 37]:

$$\sigma_{ij(E_F,T)} = q_0^2 \int_E \Xi_{ij}(E)\left(-\frac{\partial f_0}{\partial E}\right) dE, \qquad (1)$$

$$S_{ij(E_F,T)} = \frac{q_0 k_B}{\sigma_{ij}} \int_E \Xi_{ij}(E)\left(-\frac{\partial f_0}{\partial E}\right) \frac{E-E_F}{k_B T} dE, \qquad (2)$$

$$PF = \sigma S^2 \qquad (3)$$

where $\Xi_{ij}(E)$ is the Transport Distribution Function (TDF) defined below in Eq. (4), $E_F$, $T$, $q_0$, $k_B$, and $f_0$, are the Fermi level, absolute temperature, electronic charge, Boltzmann constant, and equilibrium Fermi distribution, respectively. The TDF for a simple ellipsoidal band, is given by:

$$\Xi_{(E,T)} = \frac{1}{3} v_{(E)}^2 \tau_{(E)} g_{(E)} \qquad (4)$$

where $v_{(E)} = \sqrt{\frac{E}{2m_c}}$ is the parabolic band velocity involving the conductivity effective mass $m_C$, $g_{(E)}$ the parabolic total DOS, involving the $m_{DOS}$, and $\tau(E)$ is the relaxation time. The factor $\frac{1}{3}$ picks the unidirectional velocity square component since the term $v^2$ combines partial velocities of all three cartesian orientations. In this section, for a single ellipsoidal valley, we only consider elastic scattering with acoustic phonons within the deformation potential method (ADP), which gives:

$$\frac{1}{\tau_{(E)}^{(ADP)}} = \frac{\pi}{\hbar} D_{ADP}^2 \frac{k_B T}{\rho v_S^2} g_{(E)} \qquad (5)$$

where $D_{ADP}$ is the associated deformation potential, $\rho$ is the mass density, $v_s$ the sound velocity, and $g_{(E)}$ the total density of states (DOS). We use commonly encountered values as $D_{ADP} = 5$ eV, $\rho = 6\times10^3$ Kg/m³, and $v_s = 4$ km/s. Finally, the total electrical conductivity is computed by averaging the conductivities in the three crystallographic orientations x-, y-, and z- as $\sigma_{tot} = (\sigma_{xx} + \sigma_{yy} + \sigma_{zz})/3$. The Seebeck coefficient can be computed as the weighted average of the coefficients in the different orientations, but because its orientation dependence is in this case is very weak, it was ignored.

To begin, we set an isotropic effective mass for the spherical band and compute its density of states (DOS). We then vary the longitudinal ($m_l$) and transverse ($m_t$) masses of the ellipsoidal band such that the DOS remains constant and equal to the DOS of the spherical band. In such case, we have the same number of states and perform the comparisons at the same DOS.

The $m_C$ and $m_{DOS}$ masses are given by $m_c = 3(\frac{1}{m_l} + \frac{2}{m_t})^{-1}$ and $m_{DOS} = (m_l m_t^2)^{1/3}$. $m_C$ determines the carrier velocities and $m_{DOS}$ determines the position of the Fermi level compared to the band edge, $\eta_F$, as well as the scattering rates. Thus, both quantities determine the conductivity of the material, $\sigma$. We start from the nominal case, with an isotropic spherical band as in Fig. 1a. The calculations for the electrical conductivity, $\sigma$, and PF as a function of the $\eta_F$ are shown in Fig. 2a and 2b by the lowest value lines. We then consider the anisotropic ellipsoids of the same DOS. As the anisotropy increases, essentially by increasing the ratio of the longitudinal to the transverse effective masses of the ellipsoid ($m_l/m_t$), we observe an increase in the conductivity of the material, which is translated into an increase of the PF. The Seebeck coefficient, shown in the inset of Fig. 2a, remains the same with anisotropy as it depends primarily on the distance of the $\eta_F$ from the band edge. We consider as far as $m_l/m_t = 20$, and then a limiting case of $m_l/m_t = 100$ for the very thin, ultra-elongated ellipsoidal bands, resembling thread-like energy surfaces. The latter has raised significant interest lately for the possibility of providing low-dimensional features within 3D materials [8]. The resultant $1/m_C$ varies up to values of $1/m_C = 2$ for the realistic cases and around $1/m_C = 3$ for the liming case, as shown in Fig. 2c. Note that correspondingly, even under extreme anisotropy and in the presence of ultra-stretched ellipsoids in the bandstructure (the closer one gets to threads), the ratio of $m_{DOS}/m_C$ does not vary strongly, limiting the improvements that anisotropy can provide, as also mentioned in Ref. [27]. Figure 2d shows the relative PF improvement that anisotropic bands exhibit compared to the isotropic one, which follows directly the improvement in the $m_{DOS}/m_C$ ratio, independent of $\eta_F$. We note that the same conclusions regarding the anisotropy benefits to the PF are reached when the inelastic optical phonon scattering mechanisms is also considered.

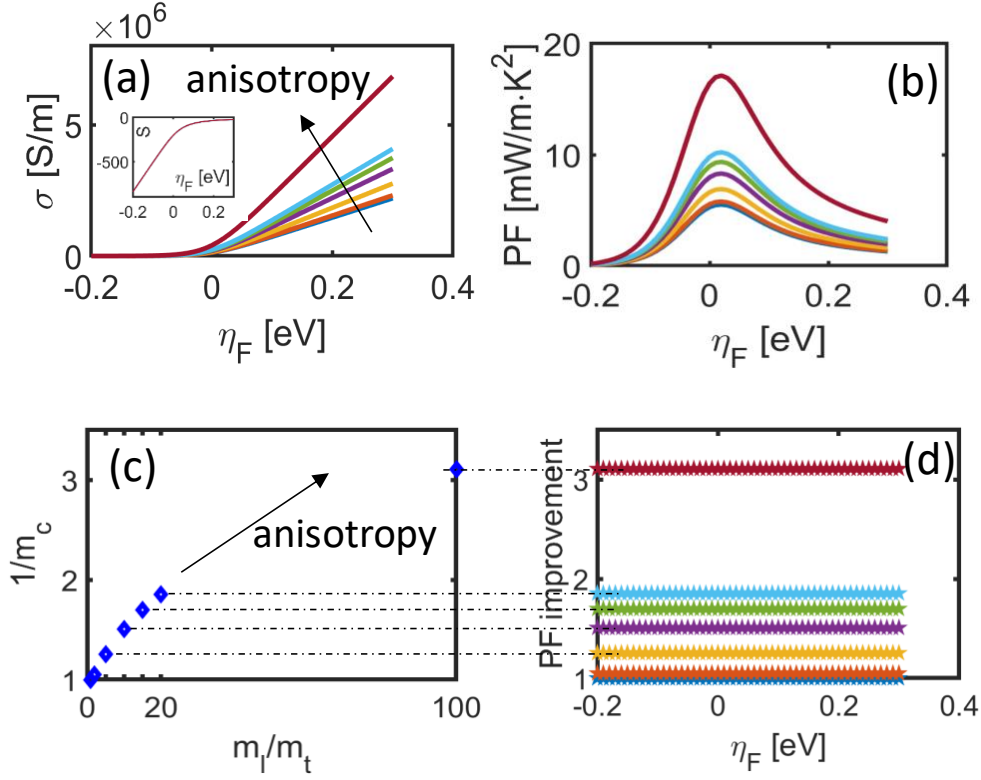

**Figure 2**: Comparison of the TE performance of a single ellipsoidal band compared to a spherical one. (a) The electrical conductivity as the ellipsoidal anisotropy increases. Inset: The Seebeck coefficient. (b) The power factor as anisotropy increases. (c) The variation of the inverse of the conductivity effective mass with the ratio of the longitudinal to transverse masses of the ellipsoid. (d) The relative PF improvement from the spherical band as anisotropy increases. The $m_l/m_t$ ratios used are: 2, 5, 10, 15, 20, 100. The explicit mass values used, in units of $m_0$, are: $m_t$ = 0.794, 0.585, 0.464, 0.405, 0.368, 0.215, $m_l$ = 1.587, 2.924, 4.642, 6.082, 7.368, 21.544. The resultant conductivity effective masses are $m_c$ = 0.952, 0.797, 0.663, 0.589, 0.539, 0.321.

## Multiple ellipsoidal bands and different scattering scenarios

We now proceed to a more realistic scenario, and consider a bandstructure as shown in Fig. 1c, in which case we assume a 6-fold degeneracy of ellipsoidal bands in the [100] equivalent crystallographic orientation. We then vary the $m_l/m_t$ ratios of the participating ellipsoids. We consider this case for two reasons: i) in any realistic bandstructure, symmetry considerations will dictate the presence of multiple ellipsoids, which average the overall energy surface into something more isotropic; ii) the presence of multiple ellipsoids dictates the consideration of inter-valley scattering in addition to intra-valley, and thus we aim to quantify the effect of band shape anisotropy under the full range of scattering mechanisms.

So far we have considered ADP-limited scattering, which is elastic and typically intra-valley. Typical TE materials consist of many valleys with inter-valley phonon scattering processes also present, primarily a result of optical phonon scattering (ODP). In addition, TE materials are highly doped to densities of ~$10^{20}$/cm$^3$, thus, ionized impurity scattering (IIS) also plays a major part in determining $\sigma$. Despite the common trend that $\sigma$ decreases with temperature in thermoelectrics, that trend is a result of a complex combination of mechanisms, and IIS is still typically a limiting factor to the magnitude of $\sigma$. As a simple example, we refer to the mobility of Si with doping, for which we have reliable data. Its value drops by almost an order of magnitude as the density increases from $10^{15}$ cm$^{-3}$ to $10^{19}$ cm$^{-3}$, indicating the well-known dominance of IIS in limiting the mobility at those doping values. At $10^{19}$ cm$^{-3}$, however, increasing the temperature still decreases the mobility, indicating that electron-phonon scattering is also strong and active in forming mobility trends. We observe the same in our typical calculations for half-Heusler materials. Furthermore, many TEs are polar materials, and polar phonon scattering (POP) is a particularly strong mechanism [38]. The latter two are anisotropic mechanisms, whereas ODP is isotropic (here scattering anisotropy refers to the dependence of the scattering rates on the exchange vector between the initial and final states, i.e their distance in the Brillouin zone). They typically allow for inter-valley transitions (albeit weakened by the distance of the initial/final scattering states) as well as intra-band. Thus, in Fig. 3 we have computed the TE coefficients for anisotropic band materials as earlier, and compare them to their isotropic band counterpart at the same DOS, but in this case we present cases for POP, ODP and IIS scattering cases.

For transport in this case, and to be able to capture the effect of anisotropic and inter-valley scattering accurately, since IIS and POP involve the exchange vector from the initial to the final scattering states, we employ our full numerical BTE solver, *ElecTra* [39], which can account for all these details, as well as arbitrary band shapes.

In this fully numerical approach, the TDF used in Eqs. 1-3, is expressed as a surface integral over the constant energy surfaces $\mathfrak{L}_E^n$, for each single band *n*, and then summed over the bands, as [37, 38, 39]:

$$\Xi_{ij(E,E_F,T)} = \frac{s}{(2\pi)^3} \sum_{k,n}^{\mathfrak{L}_E^n} v_{i(k,n)} v_{j(k,n)} \tau_{i(k,n,E_F,T)} \frac{dA_{k_{\mathfrak{L}_E^n}}}{|\vec{v}_{(k,n)}|} \quad (6)$$

where $k_{\mathfrak{L}_E^n}$ is a state on the surface $\mathfrak{L}_E^n$ and $dA_{k_{\mathfrak{L}_E^n}}$ is its corresponding surface area element. $v_{i(k,n)}$ is the *i*-component of the band velocity of the transport state, $\tau_{i(k,n)}$ is its momentum relaxation time, $\frac{dA_{k_{\mathfrak{L}_E^n}}}{|\vec{v}_{(k,n)}|}$ is its density-of-states (DOS), and *s* is the spin degeneracy. The relaxation times for each individual scattering mechanism are combined following Matthiessen's rule for each ($k, \mathfrak{L}_E^n$) state, to compute the comprehensive TE coefficients. The transport coefficients are computed along the orthogonal Cartesian space directions *x*, *y*, *z*. Consequently, the constant energy surfaces $\mathfrak{L}_E^n$ are expressed in Cartesian coordinates on orthogonal axes instead of unit cell axes, and the reciprocal unit cell is used instead of the Brillouin Zone.

For each scattering mechanism $m_s$, the corresponding momentum relaxation time $\tau_{i(k,\mathfrak{L}_E^n)}^{(m_s)}$ is defined from Fermi's golden rule as:

$$\frac{1}{\tau_{i(k,n)}^{(m_s)}} = \frac{1}{(2\pi)^3} \sum_{k'} |S_{k,k'}^{(m_s)}| \left(1 - \frac{v_{i(k',n')}}{v_{i(k,n)}}\right) \quad (7)$$

where for every initial state *k*, the sum runs over all the allowed final states *k'* of the same carrier spin [37, 38]. The $\left(1 - \frac{v_{i(k',n')}}{v_{i(k,n)}}\right)$ term is an approximation for the momentum relaxation time, which is used to solve the BTE in the closed form, as commonly done in the literature when computing the transport coefficients [40, 41]. In general, this term should contain the occupancies and relaxation times as well, as described in Ref. [42, 43], making Eq. 7 an integral equation requiring self-consistent solutions, especially for inelastic processes. However, as discussed in many places in the literature, the term we use often provides results with high accuracy, especially for low field conditions [44, 45, 46, 47], allowing for a conceptually simple evaluation of a numerically demanding integral [38].

$|S_{k,k'}|$ is the transition rate between the initial $k$ and final $k'$ states, computed for the different scattering mechanisms as:

$$\left|S_{k,k'}^{(ADP)}\right| = 2\frac{\pi}{\hbar}D_{ADP}^2\frac{k_BT}{\rho v_S^2}g_{k'} \tag{8}$$

$$\left|S_{k,k'}^{(ODP)}\right| = \frac{\pi D_{ODP}^2}{\rho\omega}\left(N_\omega + \frac{1}{2} \mp \frac{1}{2}\right)g_{k'} \tag{9}$$

$$\left|S_{k,k'}^{(POP)}\right| = \frac{\pi q_0^2 \omega}{|k-k'|^2 \varepsilon_0}\left(\frac{1}{k_\infty} - \frac{1}{k_s}\right)\left(N_\omega + \frac{1}{2} \mp \frac{1}{2}\right)g_{k'} \tag{10}$$

$$\left|S_{k,k'}^{(IIS)}\right| = \frac{2\pi}{\hbar}\frac{Z^2 q_0^4}{k_s^2 \varepsilon_0^2}\frac{N_{imp}}{\left(|k-k'|^2+\frac{1}{L_D^2}\right)^2}g_{k'} \tag{11}$$

Above, ADP is elastic scattering and ODP is inelastic. In the case of six ellipsoidal valleys, we consider ADP to be intra-valley while ODP to both intra- and inter-valley. POP is an inelastic/anisotropic scattering mechanism and is treated as both intra- and inter-valley [37]. IIS is elastic scattering, and is considered to cause both intra- and inter-valley transitions. The variables in Eqs. 8-11 are as follows: $D_{ADP}$ and $D_{ODP}$, are the deformation potentials for the ADP and ODP mechanisms; $\rho$ is the mass density; $v_s$ is the sound velocity; $\omega$ is the dominant frequency of optical phonons, considered as constant over the whole reciprocal unit cell, which has been validated to be a satisfactory approximation, [48]; $N_\omega$ is the phonon Bose-Einstein statistical distribution; $\varepsilon_0$ the vacuum permittivity, $k_s$ and $k_\infty$ the static and high frequency relative dielectric constants; $Z$ the electric charge of the ionized impurity considered; and $N_{imp}$ is the density of the ionized impurities. $g_{k'} = \frac{dA_{k'}^{n'}}{|\vec{v}_{(k',n')}|}$ is the single-spin DOS of the final scattering state. $L_D = \sqrt{\frac{k_s\varepsilon_0}{q_0}\left(\frac{\partial n}{\partial E_F}\right)^{-1}}$ is the generalized screening length with $E_F$ being the Fermi level and $n$ the carrier density [11, 37]. In the calculations performed in this section, we use the following parameters: $D_{ADP}$ = 5 eV, $D_{ODP}$ = 5×10$^{10}$ eV/m, $\rho$ = 6×10$^3$ Kg/m$^3$, $v_s$ = 4 km/s, $\hbar\omega_{ODP}$= 50 meV, $\hbar\omega_{POP}$= 40 meV, $k_s$ = 15, $k_\infty$ = 10. Note that in Eqs. 8-11 above we don't ensure energy conservation by the use of delta-functions, but by constructing energy surfaces on which all initial and final states reside. In the case of inelastic scattering, the energy difference between the initial and final energy surfaces will be the same as the energy of the phonon involved, thus energy conservation becomes numerically straightforward.

As earlier, we also extend the $m_l/m_t$ range up to 20. Figure 3a-c shows results for the TE coefficients $\sigma$, $S$, and PF under ADP scattering as before. Figure 3d shows the PF improvement ratio, which follows very similar trends as earlier, with up to a factor of 2 improvements in the PF with band anisotropy. The difference between the results in Fig. 2 and Fig. 3a-d, is that the latter case considers 6 ellipsoids rather than a single ellipsoid, which form a bandstructure which looks overall more isotropic. However, the way the conductivity is computed, i.e. by averaging the conductivities along the three different crystallographic orientations, essentially is equivalent to considering each ellipsoid independently, which justifies the equivalent results between the two figures. In addition, the PF benefits of anisotropic bands are identical in the case of the isotropic ODP inter-valley scattering alone (not show). There is no difference in the TDF when a carrier scatters into its own, or another equivalent valley, thus, inter-valley isotropic scattering processes reach the same conclusions as when considering only intra-valley processes.

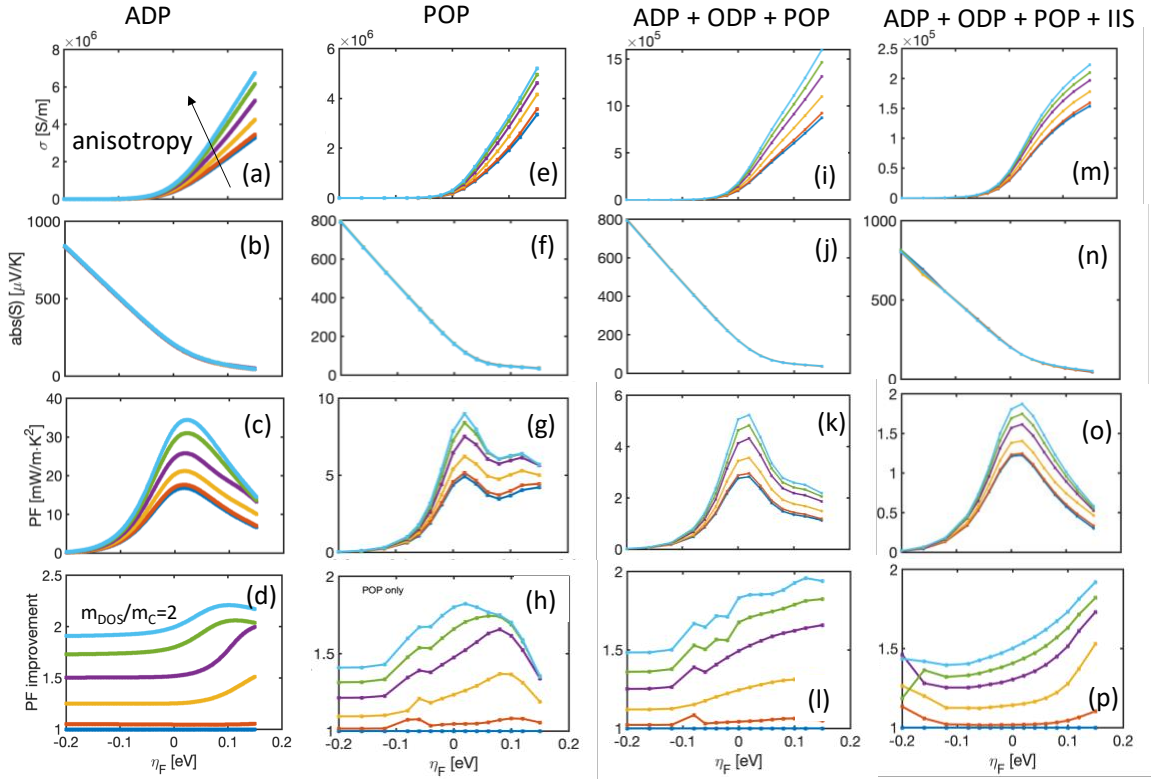

**Figure 3**: Comparison of the TE performance of a 6-fold degenerate ellipsoidal energy surface (as in Fig. 1c) compared to a spherical one. Each column shows $\sigma$, $S$, PF, and relative PF improvement as the ellipsoidal anisotropy increases. (a-d) The case of ADP

scattering only. (e-h) The case of polar optical phonon (POP) scattering. (i-l) The case of ODP+POP scattering. (m-p) The case of all ADP+ODP+POP+IIS mechanisms together.

Figure 3e-h (second column) shows the corresponding results for the case of POP scattering alone. POP is considered to be intra- and inter-valley, although the inter-valley transitions are weak. However, it is an anisotropic mechanism, and this is what we aim to examine with these simulations. The improvement of the anisotropic bands in comparison to the isotropic band persists, although it reaches somewhat lower values. For low $\eta_F$ the improvement is up to 1.5×, i.e. 50% lower, whereas at the optimal $\eta_F = 0$ eV it reaches up to 1.8× for the most elongated ellipsoids with $m_l/m_t = 20$. The reason for this reduced improvement is the anisotropic nature of POP. Anisotropic scattering mechanisms are stronger when initial and final states are nearby in $k$-space, i.e. across ellipsoidal states in the light $m_t$ transverse direction. On the other hand, they are weaker for states which are farther away, such as states across the heavy $m_l$ longitudinal direction of the ellipsoids. Thus, light states scatter more effectively, whereas heavy states scatter less, which smoothens the differences of their contribution to the TDF and makes the overall transport more isotropic. An illustration that explains this more clearly is provided in the SI, Fig. S.4.

Figures 3i-l (third column) show the corresponding results when all phonon mechanisms ADP+ODP+POP are considered. Since POP is the strongest mechanism, it dominates the trends for the PF improvements, and the values in Fig. 3l are similar to those in Fig. 3h. Note that the fact that POP is the strongest mechanism in our simulations results from our choice of deformation potentials typical of semiconductor materials. They end up providing weaker scattering rates compared to POP, which is typically a strong scattering mechanism and often dominates in polar materials [38]. Overall, improvements are still observed to the PF for anisotropic bands, albeit somewhat reduced by the anisotropic scattering nature of POP.

We now proceed in including ionized impurity scattering (IIS), another anisotropic scattering mechanism. This is an important mechanism since TE materials are typically highly doped at densities of $10^{19}$-$10^{20}$/cm$^3$. Figures 3m-p shows this most realistic case with all ADP+ODP+POP+IIS considered. Here, similarly to the last two examined cases, the addition of this extra anisotropic scattering mechanisms reduces the PF improvements even more, by ~50% to the levels of ~1.5×.

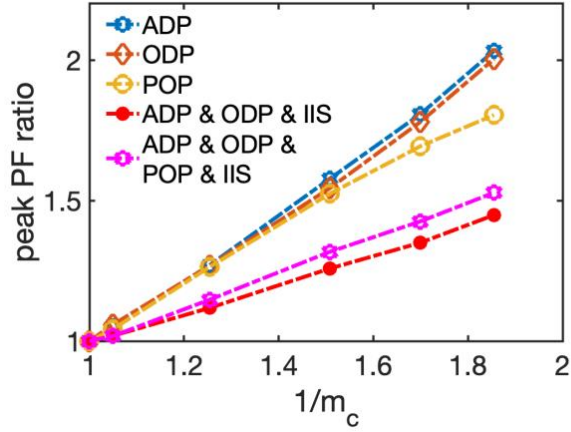

**Figure 4**: Comparison of the peak PF of a 6-fold degenerate ellipsoidal energy surface (as in Fig. 1c) compared to a spherical one, versus the inverse of the conductivity effective mass (at the same $m_{DOS}$) in units of $m_0$. Cases for different scattering mechanism considerations are shown, as indicated in the legend.

To summarize these data, in Fig. 4 we focus on the peak PF reached in the graphs of Fig. 3, and plot the PF improvement ratio versus the $1/m_C$ of the ellipsoidal bands at the specific optimal $\eta_F$ for each case. In all cases this happens at $\eta_F \sim 0$ eV as expected. Also note that the $m_C$ for the isotropic band is $m_C = 1$, thus $1/m_C = 1$ as well, i.e. the graph's origin. Considerations of solemnly ADP improve the PF of anisotropic bands by up to ~2× (far right values). The inter-valley isotropic scattering mechanism, ODP, has the same effect and still allows for a ~2× improvement. Inclusion of anisotropic scattering mechanisms such as POP and IIS reduce this improvement down to ~1.5×, which would be the more realistic case. Note that 50% in PF improvement is still quite significant for TE materials. Also note that this is the maximum realistic value for a fixed $m_{DOS}$, and it corresponds to ratios of $m_l/m_t = 20$ in ellipsoidal bands, or a $1/m_C = 2$ in more general complex shaped warped bands (the latter can also be expressed as $m_{DOS}/m_C$, which is a logical measure for anisotropy, and in this case $m_{DOS} = 1$). While such large values of $m_l/m_t$ can be rare, $m_{DOS}/m_C$ can take values more than 2 as we will see below for realistic materials.

## Anisotropy considerations for realistic DFT bands

We now proceed to examine the effect of shape anisotropy on the PF for realistic materials with DFT extracted electronic structures. For this purpose, we employ a group of 14 half-Heusler materials and we consider both their conduction and valence bands. Thus, we have two groups of 14 electronic structures for which we compute the PFs using the BTE as described above using *ElecTra* [39]. We consider the ADP, ODP and IIS scattering mechanisms, i.e we include one intra-band isotropic scattering mechanism, one intra- and inter-valley isotropic, and one intra- and inter-valley anisotropic mechanism. In this way we cover a range of mechanisms with different behavior.

The electronic transport properties depend on the scattering rates, which depend on many material specific parameters such as deformation potentials, phonon energies, sound velocities, dielectric constants, etc. These parameters vary significantly between materials. While we have access to all the necessary parameters, the purpose of this work is to isolate the effect of the band energy surface shape, rather than to quantitatively compute the PF. We have presented such a quantitative analysis on the performance of HHs and relevant comparisons elsewhere in Ref. [11, 36], also with some comparisons to experiments. Therefore, we remove the effect of all material specific parameters and extract the PF by using a common set of parameters for all materials within Eqs. 8-11 as follows: $\rho = 9$ g/cm$^3$, $v_s = 6$ km/s, ADP = 2 eV, ODP = 2x10$^{10}$ eV/m, $\hbar\omega_{\text{ODP}}= 28$ meV, which are somewhat typical for these materials [11]. In this way the PF variations between materials will be a result of the energy surface shapes alone. Thus, what we refer to as the 'PFs' of these materials are not the actual PF values, but measures of the PF only based on band shapes. Note that we do not use POP here. Since we use a common set of parameters for all materials, we essentially have relaxation times defined by a constant multiplied by some energy/momentum-dependent function. Thus, we only need to consider one mechanism for every different scattering bahavior, i.e. elastic/inelastic, isotropic/anisotropic, and intra/inter-valley. The effect of POP, which is an anisotropic and inelastic mechanism, is captured by the IIS and ODP, respectively, or by arbitrary adjustments of the constant value in the definition of the relaxation times.

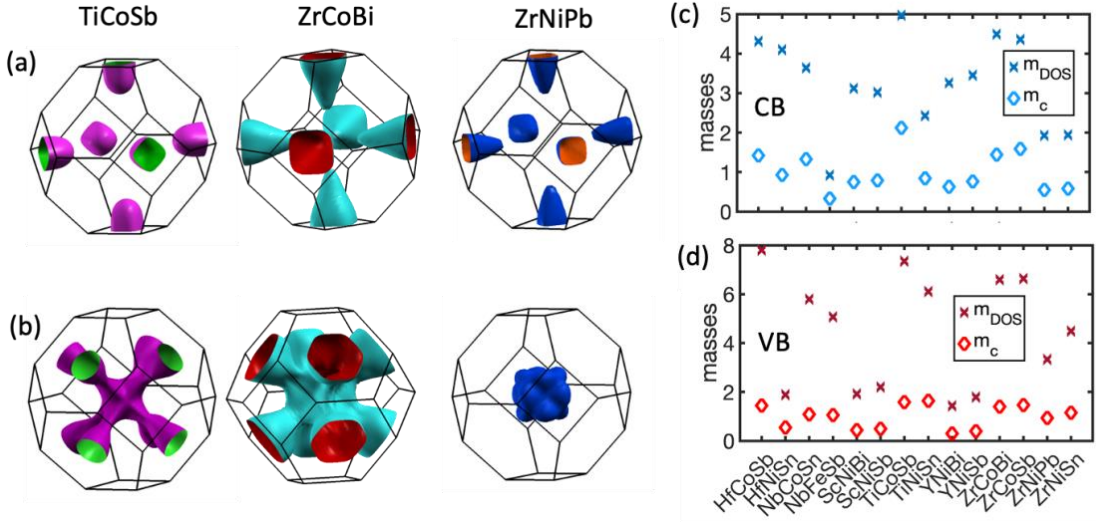

**Figure 5**: Constant energy surfaces for some half-Heusler alloys as indicated, at 0.25 eV from the band edge. (a) The conduction bands (upper row). (b) The valence bands (lower row). (c) The $m_{DOS}$ and $m_C$ for the CB. (d) The $m_{DOS}$ and $m_C$ for the VB.

In Fig. 5a-b we show band shape examples at an energy $E = 0.25$ eV into the corresponding bands, obtained via the XCrySDen package [49]. We show three energy surfaces for the conduction (Fig. 5a, upper row) and three for the valence bands (Fig. 5b, lower row). To evaluate the degree of band anisotropy and whether the calculated PF correlates with that, we need to extract the necessary values for $m_{DOS}$ and $m_C$. Since such quantities cannot be easily defined or derived for highly irregular energy surface band shapes, we proceed as follows. We compute the $m_{DOS}$ as the effective mass of an isotropic parabolic band that gives the same carrier density as the actual band structure for a specific Fermi level position in the non-degenerate regime. We evaluate $m_C$ as the effective mass of an isotropic parabolic band, which maps the average velocity of the band states weighted by their contribution to transport, again in the non-degenerate regime [11, 50]. For this, we employ a simple ballistic field effect transistor model [51], extract the average injection velocity in the sub-threshold regime, and map that velocity to a parabolic band, which provides the same subthreshold injection velocity. More details, as well as a code to extract these for arbitrary electronic structures are presented in Ref. [11, 52, 53]. These masses are, thus, ensembled quantities that can capture the various complexities of the bands such as warping and corrugation, without the need or more complex band descriptions. Figures 5c and 5d show these masses for the 14 HH materials we consider for the conduction and valence bands, respectively. With crosses we show the $m_{DOS}$ and with diamonds the $m_C$. In all cases the $m_{DOS}$ is larger than $m_C$, owing it to the high degeneracy (which is captured as well) and degree of band anisotropy.

Note that there are alternative ways to extract these masses and define anisotropy. For example, Ref. [27], utilizes Boltzmann transport to calculate the mobility of a material and its Seebeck coefficient under the constant relaxation time approximation (CRTA, $\tau_C$). The $m_C$ can be extracted from the mobility by matching it to $\mu = q_0 \tau_C/m_C$. The $m_{DOS}$ is extracted by fitting analytical calculations to the numerical Seebeck coefficient, in which case the authors extract a so-called Seebeck $m_S$, which is a measure of the $m_{DOS}$ weighted by the number of valleys [27]. In both calculations, $\tau_C$ eventually drops out of the integrals and its specific value is not of relevance. The effective number of valleys is then extracted by dividing $m_S$ by the single ellipsoid DOS mass, $m_b$, as $N^* = (m_S/m_b)^{3/2}$. The effective anisotropy parameter is given by $K^* = (m_b/m_C)^{3/2}$, which is very similar to what we consider, but the use of the 3/2 exponent points to the actual DOS rather than the mass itself (for the numerator). The surface complexity factor, which is then used as a descriptor for high performance is defined as $N^*K^* = (m_S/m_C)^{3/2}$. It is a logical method since it links the extracted masses to low-field transport properties through the use of Boltzmann transport codes (BTE). Our method does not require the use of Boltzmann transport codes, but rather extracts the $m_{DOS}$ and $m_C$ using bandstructure information alone (with the computation cost of just a few seconds). Note that although the relaxation time at first order cancels out from the ratios of the Kernel integrals that determine the transport coefficients, strictly speaking this is only allowed under the CRTA. The relaxation times of energy-, momentum-dependent (anisotropic) scattering mechanisms cannot cancel out and this can make a difference in the extracted mass values.

In Figure 6 we show a measure of the PF of these electronic structures for the $n$-type and $p$-type cases separately by blue and red symbols, respectively. Here we only pick the highest PF value from the PF vs $\eta_F$ functions (see SI for the corresponding figures). We plot these values versus variations of $m_{DOS}/m_C$ as a simple measure of the anisotropy in the bandstructure, and extract the Pearson correlation coefficient for each group. The reason we treat the two distributions separately is that the VB has large valley degeneracy, and thus its distribution appears shifted to the right towards larger $m_{DOS}/m_C$ ratios, since the effect of valley anisotropy is included in $m_{DOS}$. If we try to combine the two groups that are shifted on the x-axis, we will reach anticorrelated results with $m_{DOS}/m_C$.

Figure 6a shows the calculated peak PF versus the $m_{DOS}/m_C$. The Pearson correlation coefficient for the CB is $r_e = 0.81$ and for the VB is $r_h = 0.63$. The high correlation, especially in the case of electrons, indicates that $m_{DOS}/m_C$ can describe the degree of anisotropy in realistic electronic structures, and the PF correlates with that. Another important observation is the degree of variation in the $m_{DOS}/m_C$ and the PF. By comparing at the lowest PF values for each group, in the case of the *n*-type materials the band anisotropy $m_{DOS}/m_C$ increases by ~2.5×, and correspondingly the PF experiences a ~3× improvement (blue symbols). For *p*-type materials a factor of somewhat less than 2× is observed for the anisotropy variation, whereas a ~2.5× PF improvement is observed (red symbols). The fact that we observe larger PF improvements compared to the analytical case in Fig. 2, is because previously we have considered the same $m_{DOS}$ for all anisotropic bands and only allowed the $m_C$ to vary. Here we cannot control either $m_{DOS}$ or $m_C$, thus the variation can exceed the analytical values.

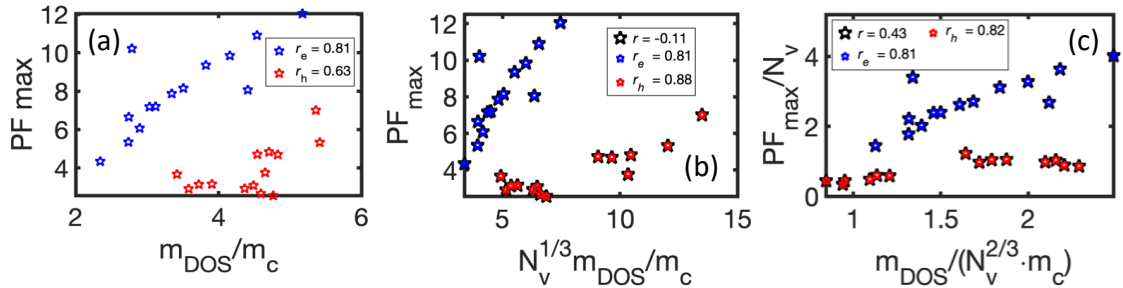

**Figure 6**: (a) Simplest way to quantify the anisotropy as variants of $m_{DOS}/m_C$, indicating high PF-measure correlation. (b) Improving the x-axis variable to fully account for the number of valleys $N_V$, which are already included in the y-axis PF. (c) Removing the effect of $N_V$ from both the x- and y-axis variables to focus on band anisotropy alone. In the insets we show the Pearson correlation coefficients for the CB (blue), the VB (red), and all combined (black).

The correlation for *p*-type materials is not as high as we would like it to be, thus below we modify the x-axis variable to improve it. In the above discussion for Fig. 6a, the $m_{DOS}$ includes the effect of valley degeneracy, $N_V$. Since the computed PF 'measure' in the y-axis involves all bandstructure valleys, we modify the x-axis variable to include that as well. Noticing that $m_{DOS}/m_C = N_V^{2/3} m_{DV}/m_C$, where $m_{DV}$ is the averaged DOS mass per valley, we then multiply by $N_V^{1/3}$ to reach $N_V^{1/3} m_{DOS}/m_C = N_V m_{DV}/m_C$, which should correlate better with the PF involving $N_V$ [54]. Indeed, this is shown in Fig. 6b, where the correlations have now reached $r_e = 0.81$ and $r_p = 0.88$. This particularly benefits the correlation for the *p*-type materials which have larger valley degeneracy, since the x-axis is now directly related to $N_V$ in the same way as $N_V$ benefits the PF in the y-axis.

Although a high degree of correlation can be achieved with $N_V^{1/3} m_{DOS}/m_C$, our focus in this work is to isolate the effect of anisotropy alone. The benefits of high $N_V$ are well documented by us and others in prior works [11, 12, 27]. Thus, in Fig. 6c we remove the effect of $N_V$ from both the PF and the $N_V^{1/3} m_{DOS}/m_C$ by dividing both by $N_V$. Thus, we plot PF/$N_V$ versus $m_{DOS}/(N_V^{2/3} m_C) = m_{DV}/m_C$. The figure shows the averaged contribution to the PF per valley versus the valley averaged anisotropy per valley, $m_{DV}/m_C$. The correlation still remains high for both *n*-type and *p*-type with $r_e = 0.81$ and $r_p = 0.82$. We note that the all CBs have their minima at the X-point, so the number of valleys is the same for all the *n*-type materials. Again, the valley averaged anisotropy varies by ~2.5× for both cases. The PF/$N_V$ improvement is of the order of ~3.5× for electrons and ~2.5× for holes. Note than in this case the two distributions are shifted around the same origin at the left side of the plot and if they are treated as one combined distribution, the correlation coefficient is increased to $r_{tot} = 0.43$, rather than negative as in the previous cases (indicated in the legend of Fig. 6b). We stress again here that the correlations presented in Fig. 6 do not reflect the PF correlation and ranking of the actual HH materials, due to the use of fixed parameters for all compounds and the isolated use of the bandstructure shape. Actual performance correlations are described by more involved descriptors that include transport parameters such as deformation potentials, as presented in Ref. [11].

Finally, we would like to provide a word of caution. Such high correlation coefficients, could provide the indication that these quantities to describe anisotropy can be used effectively as descriptors in materials screening studies to identify anisotropic band shape materials with benefits to the PF. However, this is not always the case and careful considerations are further needed. We know from our previous works that different transport formalisms lead to different descriptors in the search of high-performance materials through materials screening studies [11]. For example, the use of the CRTA, suggests that large $m_{DOS}$ will be beneficial for the PF. On the other hand, transport which considers the full energy/momentum/band-dependence of the scattering mechanisms, leads to descriptors that involve the $1/m_C$ term, but exclude the $m_{DOS}$ term [11]. The reason is that the conductivity, as determined by the TDF, involves the product of $v_{(E)}^2 \tau_{(E)} g_{(E)}$. Since the scattering times are inversely proportional to the density of states, what remains is $v_{(E)}^2$, which is proportional to $1/m_C$, whereas $m_{DOS}$ drops out of the picture at first order. Since again at first order the Seebeck coefficient depends on the distance of the Femi level from the bands, $\eta_F$, and the maximum PF typically coincides with $\eta_F = 0$ eV at first order, it turns out that the PF is determined only by $1/m_C$. Prior studies that used the CRTA suggested that a variation of $m_{DOS}/m_C$ as a measure of anisotropy, where the numerator had the "Seebeck mass" $m_S$ instead of $m_{DOS}$, correlates well with the so-calculated PF of a large number of TE materials [27]. Although in this work we also used versions of $m_{DOS}/m_C$ as a measure of anisotropy to suggest high PF materials, we would like to mention that typically the PF predictions from the CRTA and those from the energy-dependent scattering times can be anticorrelated [11].

In fact, a large $m_{DOS}$ can negatively affect the PF correlation [11]. This is also what we observe in this work if we compute the Pearson correlation coefficient for the PF with the $N_V$ weighted variants of $m_{DOS}/m_C$ for the two groups of 14 HH bandstructures, while considering only ADP or only ODP scattering mechanisms. In those cases (see Supporting Information, SI), the correlation coefficients are either very low or even negative, while high correlation is provided by $1/m_C$ alone. The high correlation in Fig. 6 is a result of the IIS anisotropic scattering mechanisms. This is because the scattering times for IIS are proportional to the exchange vector $(\boldsymbol{k} - \boldsymbol{k}')$ in addition to the inverse DOS (see Eqs. 10 and 11). Thus, the TDF(E) in this case becomes $\sim (\boldsymbol{k} - \boldsymbol{k}')^2 v_{(E)}^2$. A larger exchange vector at a given energy reflects an increased $m_{DOS}$ as well, which makes the final $m_{DOS}/m_C$ quantity reasonably fit to capture PF trends. This is the same as when using the CRTA, in which case the TDF~$v_{(E)}^2 g_{(E)}$. Note however, that as a descriptor for searching high PF materials, $1/m_C$ alone provides better results compared to $m_{DOS}/m_C$ for any (energy dependent) scattering mechanisms considered [11].

## Conclusions

In this work we have investigated the effect of the shape of the electronic bandstructure energy surfaces, and specifically their anisotropy and warping in enhancing the thermoelectric power factor. We consider both, analytical electronic bands and fully numerical DFT extracted bands, and perform electronic transport simulations using the Boltzmann transport equation including many scattering mechanisms. By isolating the band shape alone, we find that the shape of anisotropic/warped bands can result in PF improvements by up to ~3× compared to fully isotropic band shapes. This is broadly valid in the presence of isotropic scattering mechanisms, whether intra- or inter-valley, independent of the Fermi level position. Anisotropic scattering processes like polar optical phonon and ionized impurity scattering, on the other hand, can reduce these improvements by ~25-50% depending on the Fermi level position. We find that the ratio of the density of states effective mass to the conductivity effective mass ($m_{DOS}/m_C$) and some of its variants which include the number of valleys, i.e. the ratio of the DOS mass per valley over the conductivity mass per valley, $m_{DV}/m_C$, can act as anisotropy measure. Finally, we describe an appropriate and efficient way to extract $m_{DOS}$ and $m_C$ from highly irregular bandstructures, without the use of Boltzmann Transport solvers and their relevant software.

## Supporting Information

Thermoelectric charge transport coefficients for the half-Heusler alloys under investigation. Pearson correlation coefficients between maximum PF and several descriptors for selected separate scattering mechanisms. Explanation of the role of the anisotropic scattering.

## Acknowledgements

This work has received funding from the European Commission under the Grant agreement 788465 (GENESIS), the European Research Council (ERC) under the European Union's Horizon 2020 research and innovation programme (grant agreement No 678763), and from the UK Research and Innovation fund (project reference EP/X02346X/1). PG received partial funding under the National Recovery and Resilience Plan (NRRP), Mission 04 Component 2 Investment 1.5 – NextGenerationEU, Call for tender n. 3277 dated 30/12/2021, Award Number: 0001052 dated 23/06/2022.

# References

[1] Beretta, D.; Neophytou, N.; Hodges, J. M.; Kanatzidis, M. G.; Narducci, D.; Martin-Gonzalez, M.; Beekman, M.; Balke, B.; Cerretti, G.; Tremel, W.Thermoelectrics: From History, a Window to the Future. *Mater. Sci. Eng. R Rep.* **2019**, 138, 100501.

[2] Artini, C.; Pennelli, G.; Graziosi, P.; Li, Z.; Neophytou, N.; Melis, C.; Colombo, L.; Isotta, E.; Lohani, K.; Scardi, P.; Castellero, A.; Baricco, M.; Palumbo, M.; Casassa, S.; Maschio, L.; Pani, M.; Latronico, G.; Mele, P.; Di Benedetto, F.; Contento, G.; De Riccardis, M. F.; Fucci, R.; Palazzo, B.; Rizzo, A.; Demontis, V.; Prete, D.; Isram, M.; Rossella, F.; Ferrario, A.; Miozzo, A.; Boldrini, S.; Dimaggio, E.; Franzini, M.; Galliano, S.;Barolo, C.; Mardi, S.; Reale, A.; Lorenzi, B.; Narducci, D.; Trifiletti, V.; Milita, S.; Bellucci, A.; Trucchi, D. M. Roadmap on thermoelectricity. *Nanotechnology*, **2023**, 34, 292001.

[3] Pei, Y.; Wang, H.; Snyder, G. Band Engineering of Thermoelectric Materials. *Adv. Mater.* **2012**, 24, 6125.

[4] Wang, X.; Askarpour, V.; Maassen, J.; Lundstrom, M. On the calculation of Lorenz numbers for complex thermoelectric materials. *J. Appl. Phys.* **2018**, 123, 055104.

[5] Heremans, J. P.;Jovovic, V.; Toberer, E. S.; Saramat, A.; Kurosaki, K.; Charoenphakdee, A.; Yamanaka, S.; Snyder, G. J. Enhancement of thermoelectric efficiency in PbTe by distortion of the electronic density of states. *Science* **2008**, *321* (5888), 554

[6] Newnham, J.A.; Zhao, T.; Gibson, Q. D.; Manning, T. D.; Zanella, M.; Mariani, E.; Daniels, L. M.; Alaria, J.; Claridge, J. B.; Corà, F.; Rosseinsky, M. J. Band structure engineering of $Bi_4O_4SeCl_2$ for thermoelectric applications. *ACS Org. Inorg. Au* 2022, **2**, 405.

[7] Perumal, S.; Samanta, M.; Ghosh, T.; Shenoy, U. S.; Bohra, A. K.; Bhattacharya, S.; Singh, A.; Waghmareand U. V.; Biswas, K. Realization of High Thermoelectric Figure of Merit in GeTe by Complementary Co-doping of Bi and In, *Joule*, **2019**, 3, 2565.

[8] Parker, D.; Chen, X.; Singh, D. J. High three-dimensional thermoelectric performance from low-dimensional bands. *Phys. Rev. Lett.,* **2013**, 110, 146601.

[9] Müchler, L.; Casper, F.; Yan, B.; Chadov, S.; Felser, C. Topological insulators and thermoelectric materials. *Phys. Status Solidi RRL* **2013**, 7, 91.

[10] Hinterleitner, B.; Knapp, I.; Poneder, M.; Shi, Y.; Müller, H.; Eguchi, G.; Eisenmenger-Sittner, C.; Stöger-Pollach, M.; Kakefuda, Y.; Kawamoto, N. Thermoelectric Performance of a Metastable Thin-Film Heusler Alloy. *Nature* **2019**, *576*, 85

[11] Graziosi, P.; Kumarasinghe, C.; Neophytou, N. Material Descriptors for the Discovery of Efficient Thermoelectrics. *ACS Applied Energy Materials* **2020**, *3*, 5913

[12] Pei, Y.; Shi, X.; LaLonde, A.; Wang, H.; Chen, L.; Snyder, G. J. Convergence of Electronic Bands for High Performance Bulk Thermoelectrics. *Nature* **2011**, *473*.

[13] Tang, Y.; Gibbs, Z. M.; Agapito, L. A.; Li, G.; Kim, H. S.; Nardelli, M. B.; Curtarolo, S.; Snyder, G. J. Convergence of Multi-Valley Bands as the Electronic Origin of High Thermoelectric Performance in $CoSb_3$ Skutterudites. *Nat. Mater.* **2015**, *14*, 1223

[14] Gorai, P.; Stevanović, V.; Toberer, E. S. Computationally guided discovery of thermoelectric materials. *Nature Reviews Materials* **2017**, *2*, 17053.

[15] Xing, G.; Sun, J.; Li, Y.; Fan, X.; Zheng, W.; Singh, D. J. Electronic Fitness Function for Screening Semiconductors as Thermoelectric Materials. *Phys. Rev. Mater.* **2017**, *1*, 065405.

[16] Kumarasinghe, C.; Neophytou, N. Band alignment and scattering considerations for enhancing the thermoelectric power factor of complex materials: The case of Co-based half-Heusler alloys. *Phys. Rev. B* **2019**, *99*, 195202.

[17] Hao, S.; Dravid, V. P.; Kanatzidis, M. G.; Wolverton, C. Computational Strategies for Design and Discovery of Nanostructured Thermoelectrics. *npj Comput. Mater.* **2019**, *5*, 1.

[18] Zhang, J.; Song, L.; Madsen, G. K.; Fischer, K. F.; Zhang, W.; Shi, X.; Iversen, B. B. Designing High-Performance Layered Thermoelectric Materials Through Orbital Engineering. *Nat. Commun.* **2016**, *7*, 10892.

[19] Hung, N. T.; Nugraha, A. R.; Yang, T.; Zhang, Z.; Saito, R. Thermoelectric performance of monolayer InSe improved by convergence of multivalley bands. *J. Appl. Phys.* **2019**, *125*, 082502.

[20] Srinivasan, B.; Gellé, A.; Gucci, F.; Boussard-Pledel, C.; Fontaine, B.; Gautier, R.; Halet, J.-F.; Reece, M. J.; Bureau, B. Realizing a Stable High Thermoelectric $zT \sim 2$ over a Broad Temperature Range in $Ge_{1-x-y}Ga_xSb_yTe$ via Band Engineering and Hybrid Flash-SPS Processing. *Inorg. Chem. Front.* **2019**, *6*, 63.

[21] Zhang, L. J.; Qin, P.; Han, C.; Wang, J.; Ge, Z.-H.; Sun, Q.; Cheng, Z. X.; Li, Z.; Dou, S. X. Enhanced Thermoelectric Performance through Synergy of Resonance Levels and Valence Band Convergence via Q/In (Q = Mg, Ag, Bi) Co-Doping. *J. Mater. Chem. A* **2018**, *6*, 2507.

[22] Chung, D.-Y.; Hogan, T.; Brazis, P.; Rocci-Lane, M.; Kannewurf, C.; Bastea, M.; Uher, C.; Kanatzidis, M. G. $CsBi_4Te_6$: A High-Performance Thermoelectric Material for Low-Temperature Applications. *Science* **2000**, 287, 1024.

[23] Mishra, S. K.; Satpathy, S.; Jepsen, O. Electronic Structure and Thermoelectric Properties of Bismuth Telluride and Bismuth Selenide J. Phys.: Condens. Matter **1997**, 9, 461.

[24] Wang, Y.; Chen, X.; Cui, T.; Niu, Y.; Wang, Y.; Wang, M.; Ma, Y.; Zou, G. Enhanced Thermoelectric Performance of PbTe Within the Orthorhombic Pnma Phase. *Phys. Rev. B: Condens. Matter Mater. Phys.* **2007**, *76*, 155127.

[25] Kuroki, K.; Arita, R. ″Pudding Mold″ Band Drives Large Thermopower in $Na_xCoO_2$ J. Phys. Soc. Jpn. **2007**, 76, 083707.

[26] Ochi M.; Kuroki, K. Comparative First-Principles Study of Antiperovskite Oxides and Nitrides as Thermoelectric Material: Multiple Dirac Cones, Low-Dimensional Band Dispersion, and High Valley Degeneracy. *Phys. Rev. Applied* **2019**, 12, 034009.

[27] Gibbs, Z. M.; Ricci, F.; Li, G.; Zhu, H.; Persson, K.; Ceder, G.; Hautier, G.; Jain, A.; Snyder, G. J. Effective Mass and Fermi Surface Complexity Factor from Ab initio Band Structure Calculations. *npj Comput. Mater.* **2017**, *3*, 8.

[28] Chen, X.; Parker, D.; Singh, D. J. Importance of Non-Parabolic Band Effects in the Thermoelectric Properties of Semiconductors. *Sci. Rep.* **2013**, *3*, 3168.

[29] Parker, D. S.; May, A. F.; Singh, D. J. Benefits of Carrier-Pocket Anisotropy to Thermoelectric Performance: The Case of p-Type $AgBiSe_2$. *Phys. Rev. Appl.* **2015**, *3*, 064003.

[30] Zhang, J.; Iversen, B. B. Fermi Surface Complexity, Effective Mass, and Conduction Band Alignment in n-Type Thermoelectric $Mg_3Sb_{2-x}Bi_x$ from First Principles Calculations. *J. Appl. Phys.* **2019**, *126*, 085104.

[31] Mecholsky, N. A.; Resca, L.; Pegg, I. L.; Fornari, M. Theory of Band Warping and its Effects on Thermoelectronic Transport Properties Phys. Rev. B: Condens. Matter Mater. Phys. **2014**, 89, 155131.

[32] Zheng, C.; Li, J.; Wang, G.; Wang, S.; Li, J.; Zhao, H.; Zang, H.; Zhang, Y.; Zhang, Y.; Yao, J. Fine Manipulation of Terahertz Waves via All-Silicon Metasurface with an Independent Amplitude and Phase. *Nanoscale* **2021**, *13*, 5809.

[33] Usui, H.; Suzuki, K.; Kuroki, K.; Nakano, S.; Kudo, K.; Nohara, M. Large Seebeck Effect in Electron-Doped $FeAs_2$ Driven by A Quasi-One-Dimensional Pudding-Mold-Type Band Phys. Rev. B: Condens. Matter Mater. Phys. **2013**, 88, 075140.

[34] Isaacs, E. B.; Wolverton, C. Remarkable thermoelectric performance in $BaPdS_2$ via pudding-mold band structure, band convergence, and ultralow lattice thermal conductivity. *Phys. Rev. Mater.* **2019**, *3*, 015403.

[35] Li, A.; Hu, C.; He, B.; Yao, M.; Fu, C.; Wang, Y.; Zhao, X.; Felser, C.; Zhu, T. Demonstration of valley anisotropy utilized to enhance the thermoelectric power factor. *Nat. Commun.* **2021**, *12*, 5408.

[36] Graziosi, P.; Kumarasinghe, C.; Neophytou, N. Impact of the scattering physics on the power factor of complex thermoelectric materials. *J. Appl. Phys.* **2019**, *126*, 155701.

[37] Nag, B. R. Electron Transport in Compound Semiconductors. Springer, Berlin, Heidelberg: vol. 11, 1981.

[38] Lundstrom, M. Fundamentals of Carrier Transport. Cambridge University Press (2000).

[39] Graziosi, P.; Li, Z.; Neophytou ,N. ElecTra code: Full-band electronic transport properties of materials. *Comput. Phys. Comm.* **2023**, 287 ,108670

[40] Sohier, T.; Campi, D.; Marzari, N.; Gibertini, M. Mobility of Two-Dimensional Materials from First Principles in An Accurate and Automated Framework. *Phys. Rev. Mater.* **2018**, *2*, 114010 .

[41] Neophytou, N.; Kosina, H. Phys. Rev. B **2011**, 84, 085313.

[42] Fischetti, M. V. J. Appl. Phys. **2001**, 89, 1232.

[43] Fischetti, M. V.; Ren, Z.; Solomon, P. M.; Yang, M.; Rim, K. J. Appl. Phys. **2003**, 94, 1079.

[44] Buin, A. K., Verma, A., Svizhenko, A., and Anantram, M. P. Nano Lett. **2008**, 8, 760.

[45] Buin, A. K.; Verma, A.; Anantram, M. P. J. Appl. Phys. **2008**, 104, 053716.

[46] Ramayya, E. B.; Vasileska, D.; Goodnick, S. M.; Knezevic, I. J. Appl. Phys. **2008**, 104, 063711.

[47] Jin, S.; Fischetti, M. V.; Tang, T. Modeling of Electron Mobility in Gated Silicon Nanowires at Room Temperature: Surface Roughness Scattering, Dielectric Screening, and Band Nonparabolicity. *J. Appl. Phys.* **2007**, *102*, 083715.

[48] Samsonidze, G.; Kozinsky, B. Accelerated Screening of Thermoelectric Materials by First-Principles Computations of Electron-Phonon Scattering. *Adv. Energy Mater.* **2018**, *8*, 1800246.

[49] Kokalj, A. XCrySDen—A New Program for Displaying Crystalline Structures and Electron Densities. *J. Mol. Graphics Modell.* **1999**, *17*, 176.

[50] Neophytou, N.; Kosina, H. Large Enhancement in Hole Velocity and Mobility in *p*-Type [110] and [111] Silicon Nanowires by Cross Section Scaling: An Atomistic Analysis. *Nano Lett.* **2010**, *10*, 4913.

[51] Rahman, A.; Guo, J.; Datta, S.; Lundstrom, M. Theory of Ballistic Nanotransistors. *IEEE Trans. Electron Devices* **2003**, *50*, 1853.

[52] Li, Z.; Graziosi, P.; Neophytou, N. Electron and Hole Mobility of SnO$_2$ from Full-Band Electron–Phonon and Ionized Impurity Scattering Computations *Crystals* **2022**, 12, 1591.

[53] Graziosi, P.; Neophytou, N. **2019**, https://doi.org/10.48550/arXiv.1912.10924 and https://github.com/PatrizioGraziosi/EMAF-code

[54] The product of the number of valleys, $N_V$, times the DOS mass *per valley*, $m_{DV}$, is $N_V m_{DV}/m_C$. In general, we don't have access to $m_{DV}$, but only to the total DOS mass, $m_{DOS}$, which contains the number of valleys as well. Thus, we need to substitute the equivalent of $m_{DV}$. By looking at the standard textbook DOS formula , it holds that: .

Thus,

"For Table of Contents Use Only."

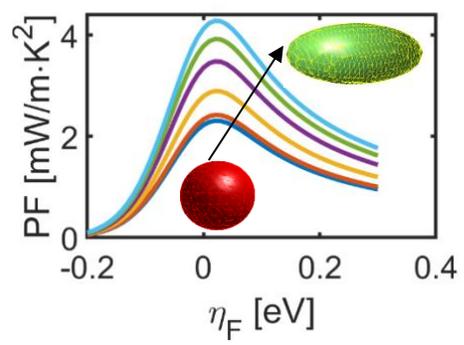